\documentclass[aps,pre,twocolumn,superscriptaddress,longbibliography,nobibnotes,nodoi,nourl,noeprint]{revtex4-2}
\usepackage{amsmath,amssymb,physics,bm,siunitx}
\usepackage{xcolor,graphicx,soul}
\usepackage[colorlinks=true,citecolor=blue,urlcolor=blue,linkcolor=black]{hyperref}

\definecolor{darkred}{rgb}{0.6, 0.2, 0.2}

\begin{document}

\title{Current reversals in driven lattice gases and Brownian motion}

\author{Moritz Wolf}
\affiliation{Universit\"at Osnabr\"uck, Fachbereich
  Mathematik/Informatik/Physik, Barbarastra{\ss}e 7, D-49076
  Osnabr\"uck, Germany}

\author{S\"oren Schweers} \email{sschweers@uos.de}
\affiliation{Universit\"at Osnabr\"uck, Fachbereich
  Mathematik/Informatik/Physik, Barbarastra{\ss}e 7, D-49076
  Osnabr\"uck, Germany}

\author{Philipp Maass} \email{maass@uos.de}
\affiliation{Universit\"at Osnabr\"uck, Fachbereich
  Mathematik/Informatik/Physik, Barbarastra{\ss}e 7, D-49076
  Osnabr\"uck, Germany}

\date{February 9, 2026}

\begin{abstract}
Particle currents flowing against an external driving are a
fascinating phenomenon in both single-particle and interacting
many-particle systems. Underlying physical mechanisms of such current
reversals are not fully understood yet. Predicting their appearance
is difficult, in particular for interaction-induced ones that emerge
upon changes of the particle density.  We here derive conditions on
external time-dependent drivings, under which current reversals occur
in lattice gases with arbitrary pair interactions. Our derivation is
based on particle-hole symmetry and shows that current reversals must
emerge if the time-varying driving potential changes sign after a
translation in time and/or space. Our treatment includes nonstationary
dynamics and time-dependent spatially averaged currents in
nonequilibrium steady states. It gives insight also into possible
occurrences of current reversals in continuous-space dynamics, which
we demonstrate for hardcore interacting particles driven across a
periodic potential by a traveling wave.
\end{abstract}

\maketitle

\section{Introduction}
\label{sec:introduction}
Reversals of particle currents have frequently been reported as
changes of current sign when varying parameters of driving forces.
They were theoretically studied for a large variety of systems
\cite{Jung/etal:1996, Mateos:2000, Kostur/Luczka:2001, Jain/etal:2007,
  Marathe/etal:2008, daSilva/etal:2008, Vincent/etal:2010,
  Zhu/etal:2016, Li/etal:2016a, Li/etal:2017, Maggi/etal:2018,
  Chen/etal:2019, Chen/etal:2022}, and observed experimentally in a
nanofluidic rocking Brownian motor \cite{Schwemmer/etal:2018}.

More specifically, one speaks about a reversal of current when it
flows against an external driving.  Particles in corresponding systems
can be viewed as having a negative effective mobility
\cite{Ros/etal:2005}. These negative mobilities were found in systems
with periodically arranged obstacles for single-particle overdamped Brownian
motion in a channel under time-periodic forcing
\cite{Eichhorn/etal:2002a, Eichhorn/etal:2002b} and for single-particle jump motion with
time-independent bias in a three-layer lattice model
\cite{Cleuren/VandenBroeck:2003}. More recently, 
they were reported for  underdamped Brownian motion of a single particle in
periodic potentials under time-periodic driving
in regimes of both normal and anomalous diffusion 
\cite{Slapik/etal:2019, Spiechowicz/etal:2019, Wisniewski/Spiechowicz:2022}, as well as
for non-Markovian baths described
by the generalized Langevin equation \cite{Wisniewski/Spiechowicz:2024, Wisniewski/Spiechowicz:2025}.
They can occur also for passive, time-independent driving, as shown for the tracer mobility in a model
of interacting particles suspended in a chiral fluid \cite{Kalz/etal:2025}.

Experimentally, negative mobilities have been observed
in the current-voltage characteristics of a resistively and
capacitively shunted Josephson junction when applying both a dc and ac
current \cite{Nagel/etal:2008}. Such experiments can be related to
particle transport, because the phase difference between the two
superconducting wave functions and hence the time-integrated voltage
across the junction evolves in time like a driven pendulum with
damping \cite{Stewart:1968, McCumber:1968}, or, when taking into
account noise, a particle performing underdamped Brownian motion in a
tilted washboard potential under time-periodic forcing
\cite{Coffey/Kalmykov:2011-chap5}.

Particularly intriguing are current flows against external driving
caused by particle interactions \cite{deSouzaSilva/etal:2006,
  Slanina:2009a, Slanina:2009b, Rana/etal:2018, Das/etal:2023}. These
reversals typically occur when the particle density exceeds a
threshold. Changes of signs in current-density relations have been
analyzed in the past for hardcore interacting Brownian particles
driven across an asymmetric sawtooth potential by a sinusoidal wave or
alternating force \cite{Derenyi/Vicsek:1995}, and a dichotomous
flashing of the sawtooth ratchet \cite{Derenyi/Ajdari:1996}. 
Motions against an external driving were furthermore observed
for vortices in superconductors due to the interplay of pinning potentials and 
vortex-vortex interactions \cite{Villegas/etal:2003, Dinis/etal:2007}.

Interaction-induced current reversals occur also in lattice gases
driven by traveling waves \cite{Chaudhuri/Dhar:2011,
  Chatterjee/etal:2014, Dierl/etal:2014, Chaudhuri:2015,
  Chatterjee/etal:2016}.  For such lattice gases, valuable insight
into underlying physical mechanisms was obtained by perturbative
calculations \cite{Chaudhuri/Dhar:2011, Chaudhuri:2015} and analytical
treatments of moving defect sites representing traveling-wave
potentials \cite{Chatterjee/etal:2014, Chatterjee/etal:2016}.  In
nonequilibrium steady states, time-averaged currents $\bar J(\rho)$ as
a function of the particle density $\rho$ were frequently found to
exhibit the symmetry property $\bar J(\rho)=-\bar J(1-\rho)$.  This
suggests that they can be understood from particle-hole symmetry
\cite{Chaudhuri/Dhar:2011, Dierl/etal:2014}. 

Other types of current reversals in lattice gases were reported
for models of molecular motors with different polarities \cite{Muhuri/Pagonabarraga:2010} 
and driven transport in a burnt-bridge exclusion process, where
particle hops actively modify the link properties between the lattice sites \cite{Schulz/etal:2011}.
In the first case, the key mechanism behind the reversal lies in the interconversion of the different motor species.
In the second case, the key mechanism is a memory effect due to nonlocal trail-mediated interactions.

Here we derive general conditions for interaction-induced current
reversals in driven lattice gases based on particle-hole
symmetry. These conditions are applicable for arbitrary pair
interactions and state that a sign inversion of a time-varying driving
potential upon a translation in space and/or time leads to
point-symmetric reversals $\bar J(\rho)=-\bar J(1\!-\!\rho)$ of
time-averaged currents in nonequilibrium steady states. 
Spatially averaged
currents in time-periodic steady states can depend on time~$t$. For such
currents $\bar J(t,\rho)$, we show that $\bar J(t,\rho)=-\bar
J(t,1-\rho)$, if the time is measured in the same frame of the
external driving.  In nonstationary dynamics, currents for
particle-hole inverted initial conditions are related.

For overdamped Brownian motion of hardcore interacting particles in a
flat potential, interaction-induced current reversals do not occur
under driving by traveling waves \cite{Chaudhuri/etal:2015}. This can
be understood from an analysis of entropy production
\cite{Lips/etal:2019}.  Using insights from our derivations for
lattice gases under time-dependent driving, we show that such
reversals appear when the driven motion takes place in a periodic
potential.

Current reversals upon a change of particle density are also possible
without time-dependent external driving.  For lattice gases, they were
reported for heterogeneous exchange rates of different particle types
in multispecies lattice gases \cite{Chatterjee/Mohanty:2018}, and
recently for generalized forms of hopping rates in the
Katz-Lebowitz-Spohn model \cite{Ngoc/Schuetz:2025, Ngoc/Nhung:2025}.

\section{Lattice gases driven by time-varying external potential}
\label{sec:TW-driven-lattice-gas-model}
We consider a one-dimensional lattice with lattice constant $a=1$ and
$L$ sites $i=1,\ldots, L$, where each site $i$ can be occupied by at
most one particle. The microstates of the system are given by sets
$\bm n=\{n_1,\ldots,n_L\}$ of occupation numbers
\begin{equation}
n_i=\left\{\begin{array}{ll}
1\,, & \mbox{if site is occupied,}\\[0.5ex]
0\,, & \mbox{otherwise.}
\end{array}\right.
\end{equation}
The number of particles is $N=\sum_{i=1}^L n_i$ and the particle density
\begin{equation}
\rho=\frac{N}{L}=\frac{1}{L}\sum_{i=1}^L n_i\,.
\end{equation}
The lattice gas Hamiltonian in the absence of driving is
\begin{equation}
H_0(\bm n)=\sum_{\alpha}\sum_{i=1}^L V_\alpha n_in_{i+\alpha}\,,
\label{eq:H0}
\end{equation}
where $V_\alpha$ are pair interactions and we
assume periodic boundary conditions, $n_{i+L}=n_i$.

An external potential
\begin{equation}
U[\bm n,\bm\epsilon(t)]=\sum_{i=1}^L \epsilon_i(t)n_i\,,
\end{equation}
drives the particles, where
$\bm\epsilon(t)=\{\epsilon_1(t),\ldots,\epsilon_L(t)\}$ is a set of
site energies varying in time $t$.  The time-dependent Hamiltonian is
\begin{equation}
H[\bm n,\bm\epsilon(t)]=H_0(\bm n)+U[\bm n,\bm\epsilon(t)]\,.
\end{equation}

The particles can jump to vacant nearest-neighbor sites, where the
jump rates $\Gamma_i^-[\bm n,\bm\epsilon(t)]$ and $\Gamma_i^+[\bm
  n,\bm\epsilon(t)]$ from site $i$ to the left and right depend at all
times on the difference between energies of the final and initial
microstate after and before the jump. For a general specification of
such rates, we introduce
\begin{equation}
\Delta H_{i,i+1}[\bm n,\bm\epsilon(t)]
=H[\bm n^{(i,i+1)},\bm\epsilon(t)]-H[\bm n,\bm\epsilon(t)]\,,
\label{eq:DeltaH}
\end{equation}
that is the difference between
the energies of a
microstate $\bm n$ and the same state $\bm n^{(i,i+1)}$ with occupation
numbers $n_i$ and $n_{i+1}$ interchanged,
\begin{align}
n_j^{(i,i+1)}=\left\{\begin{array}{cl}
n_j\,, & j\ne i,i\!+\!1\,,\\
n_i\,,  & j=i\!+\!1\,,\\
n_{i+1}\,, & j=i\,.
\end{array}\right.
\end{align}
It holds $\bm n^{(i,i+1)}=\bm n^{(i+1,i)}$ and 
$\Delta H_{i,i+1}=\Delta H_{i+1,i}$.

With this notation, the jump rates can be written as
\begin{align}
\Gamma_i^+[\bm n,\bm\epsilon(t)]&=n_i\tilde n_{i+1}f(\Delta H_{i,i+1}[\bm n,\bm\epsilon(t)])\,,
\label{eq:DefGamma+}\\[0.5ex]
\Gamma_i^-[\bm n,\bm\epsilon(t)]&=n_i\tilde n_{i-1}f(\Delta H_{i-1,i}[\bm n,\bm\epsilon(t)])\,,
\label{eq:DefGamma-}
\end{align}
where $f(.)$ gives the dependence on the energy difference, and
\begin{equation}
\tilde n_i=1-n_i
\end{equation}
is the hole occupation number. With respect to occupation numbers, we
use the tilde in an operational sense, i.e.\ $\tilde{\tilde n}_i=n_i$.

\section{Current reversals due to particle-hole symmetry}
\label{sec:reasoning_current_reversal}
Using particle-hole symmetry, we first show that particle hopping
dynamics is equal to hole hopping dynamics in a system, where both
particles and holes are interchanged, and the sign of site energies
inverted.

The particle-hole symmetry of the time-dependent lattice gas Hamiltonian
\begin{equation}
H[\bm n,\bm\epsilon(t)]=H_0(\bm n)+U[\bm n,\bm\epsilon(t)]
\end{equation}
manifests itself in the following property: under inverting the sign of $\bm\epsilon(t)$
and exchanging particles by holes it reproduces itself up to a time-dependent constant $C(N,t)$,
\begin{align}
H[\tilde{\bm n},-\bm\epsilon(t)]&=H[\bm n,\bm\epsilon(t)]-C(N,t)\,,\\
C(N,t)&=(2N\!-\!L)\sum_{\alpha} V_\alpha+\sum_{i=1}^L\epsilon_i(t)\nonumber\\
&=N\,\frac{2\rho\!-\!1}{\rho}\sum_{\alpha} V_\alpha+\sum_{i=1}^L\epsilon_i(t)\,.
\end{align}
This implies
\begin{align}
\Delta H_{i,i+1}[\tilde{\bm n},-\bm\epsilon(t)]
&=H[\tilde{\bm n}^{(i,i+1)},-\bm\epsilon(t)]-H[\tilde{\bm n},-\bm\epsilon(t)]
\nonumber\\
&=\Delta H_{i,i+1}[\bm n,\bm\epsilon(t)]\,.
\label{eq:DeltaH-symmetry}
\end{align}

The dynamics can be alternatively described by nearest-neighbor
hopping of holes with rates
\begin{equation}
\tilde\Gamma_i^\pm[\bm n,\bm\epsilon(t)]=\Gamma_{i\pm1}^\mp[\bm n,\bm\epsilon(t)]\,.
\end{equation}
As a consequence of Eq.~\eqref{eq:DeltaH-symmetry}, we can relate
particle hopping rates to hole hopping rates in a system with
particles and holes interchanged:
\begin{align}
\tilde\Gamma_i^+[\tilde{\bm n},-\bm\epsilon(t)]&=\Gamma_{i+1}^-[\tilde{\bm n},-\bm\epsilon(t)]
\label{eq:relation_Gamma+}\\
&=\tilde n_{i+1}\tilde{\tilde n}_i f(\Delta H_{i,i+1}[\tilde{\bm n},-\bm\epsilon(t)])\nonumber\\[0.5ex]
&=n_i\tilde n_{i+1}f(\Delta H_{i,i+1}[\bm n,\bm\epsilon(t)])=\Gamma_i^+[\bm n,\bm\epsilon(t)]\,,
\nonumber\\[1ex]
\tilde\Gamma_i^-[\tilde{\bm n},-\bm\epsilon(t)]&=\Gamma_{i-1}^+[\tilde{\bm n},-\bm\epsilon(t)]
\label{eq:relation_Gamma-}\\
&=\tilde n_{i-1}\tilde{\tilde n}_i f(\Delta H_{i-1,i}[\tilde{\bm n},-\bm\epsilon(t)])\nonumber\\[0.5ex]
&=n_i\tilde n_{i-1}f(\Delta H_{i-1,i}[\bm n,\bm\epsilon(t)])=\Gamma_i^-[\bm n,\bm\epsilon(t)]\,.
\nonumber
\end{align}

Equations~\eqref{eq:relation_Gamma+} and \eqref{eq:relation_Gamma-}
can be utilized to determine classes of external drivings giving rise
to current reversals. They occur if the external driving potential
satisfies the following current-reversal conditions
\begin{list}{}{\setlength{\leftmargin}{1.5em}\setlength{\rightmargin}{0em}
\setlength{\itemsep}{0ex}\setlength{\topsep}{1ex}}

\item[(i)] translations in space and/or time lead to a sign inversion,

\item[(ii)\,] the translation symmetry must not give rise to a zero current.

\end{list}

We first consider sign inversion under a time translation,
\begin{equation}
\epsilon_i(t+\tau/2)=-\epsilon_i(t)\,, \hspace{1em}i=1,\ldots,L\,,
\label{eq:eps-inversion-timeshift}
\end{equation}
which implies that the site energies are $\tau$-periodic functions in
time also, $\epsilon_i(t+\tau)=\epsilon_i(t)$.  The symmetry property
\eqref{eq:eps-inversion-timeshift} together with
Eqs.~\eqref{eq:relation_Gamma+} and \eqref{eq:relation_Gamma-} yields
\begin{equation}
\tilde\Gamma_i^\pm[\tilde{\bm n},\bm\epsilon(t\!+\!\tau/2)]=
\tilde\Gamma_i^\pm[\tilde{\bm n},-\bm\epsilon(t)]=\Gamma_i^\pm[\bm n,\bm\epsilon(t)]\,.
\end{equation}
This means that hopping dynamics in a system with particles and holes interchanged
is the same after a time-translation by $\tau/2$. 

More precisely, given an initial state $\bm n^{(0)}=\bm n^{(0)}(0)$ at
a time $t=0$, a path
\begin{equation}
\mathcal{P}=\{\bm n^{(0)}(0), \bm n^{(1)}(t_1),  \bm n^{(2)}(t_2),\ldots\}
\label{eq:pathP}
\end{equation}
with subsequent jump events at times $t_1, t_2, \ldots$
in a system $\mathcal{S}$ with $N$ particles occurs with the same probability as the 
path 
\begin{equation}
\tilde{\mathcal{P}}=\{\tilde{\bm n}^{(0)}(\tau/2), \tilde{\bm n}^{(1)}(t_1+\tau/2), \tilde{\bm n}^{(2)}(t_2+\tau/2),\ldots\}
\end{equation}
in a system $\tilde{\mathcal{S}}$ with $N$ holes ($L\!-\!N$ particles) and initial
condition $\tilde{\bm n}^{(0)}$ at time $\tau/2$,
\begin{equation}
{\rm Prob}(\mathcal{P})={\rm Prob}(\tilde{\mathcal{P}})\,.
\label{eq:equal-path-prob}
\end{equation}

The particle current along the link $(i,i\!+\!1)$ from site $i$ to
$i\!+\!1$ at time $t$ in system $\mathcal{S}$ is
\begin{equation}
J_{i,i+1}(t)\Bigr|_{\mathcal{S}}
=\langle \Gamma_i^+[\bm n,\bm\epsilon(t)]-\Gamma_{i+1}^-[\bm n,\bm\epsilon(t)]\rangle\,,
\label{eq:J-in-S}
\end{equation}
where the average $\langle\ldots\rangle$ is over all paths for given initial
state $\bm n^{(0)}$. Likewise, the link current of holes is
\begin{align}
\tilde J_{i,i+1}(t)\Bigr|_{\mathcal{S}}
&=\langle \tilde\Gamma_i^+[\bm n,\bm\epsilon(t)]-\tilde\Gamma_{i+1}^-[\bm n,\bm\epsilon(t)]\rangle\nonumber\\
&=\langle \Gamma_{i+1}^-[\bm n,\bm\epsilon(t)]-\Gamma_i^+[\bm n,\bm\epsilon(t)]\rangle\,.
\end{align}
Because of Eq.~\eqref{eq:equal-path-prob}, the particle current in
Eq.~\eqref{eq:J-in-S} must equal the hole current in system
$\tilde{\mathcal{S}}$ at time $t+\tau/2$.  Moreover, hole currents are
the negative of particle currents. Hence,
\begin{equation}
J_{i,i+1}(t)\Bigr|_{\mathcal{S}}=\tilde J_{i,i+1}(t\!+\!\tau/2)\Bigr|_{\tilde{\mathcal{S}}}
=-J_{i,i+1}(t\!+\!\tau/2)\Bigr|_{\tilde{\mathcal{S}}}.
\label{eq:J-relation-nonstationary-1}
\end{equation}
This relation shows that current reversals can even be identified in
time-shifted nonstationary dynamics when having particle-hole
interchanged initial conditions in two systems with densities $\rho$
and $1-\rho$.

Of more practical interest are systems in time-periodic steady states, where
initial conditions are irrelevant.  For the system $\mathcal{S}$ with
particle density $\rho$, the time-averaged current is
\begin{equation}
\bar J(\rho)=\frac{1}{\tau}
\int\limits_t^{t+\tau}\!\dd t'\, J_{i,i+1}(t')\Bigr|_{\mathcal{S}}\,.
\end{equation}
The time average of $J_{i,i+1}$ in the time-periodic steady state must be
independent of $i$ due to Kirchhoff's law.  Because of
Eq.~\eqref{eq:J-relation-nonstationary-1},
\begin{align}
\bar J(1\!-\!\rho)&=\frac{1}{\tau}\hspace{-0.25em}
\int\limits_t^{t+\tau}\hspace{-0.25em}\dd t' J_{i,i+1}(t')\Bigr|_{\tilde{\mathcal{S}}}
=\frac{1}{\tau}\hspace{-0.25em}
\int\limits_{t-\tau/2}^{t+\tau/2}\hspace{-0.5em}\dd t' J_{i,i+1}(t'\!+\!\tau/2)\Bigr|_{\tilde{\mathcal{S}}}\nonumber\\
&=-\frac{1}{\tau}\int\limits_{t-\tau/2}^{t+\tau/2}\hspace{-0.5em}\dd t'\, J_{i,i+1}(t')\Bigr|_{\mathcal{S}}
=-\bar J(\rho)\,.
\label{eq:Jrelation-time-averaged}
\end{align}

For a sign inversion under a spatial translation of $m$ lattice
spacings,
\begin{equation}
\epsilon_{i+m}(t)=-\epsilon_i(t)\,, \hspace{1em}i=1,\ldots,L\,,
\label{eq:eps-inversion-spaceshift}
\end{equation}
implying $\epsilon(t)$ to be $2m$-periodic in the lattice, the
reasoning for the occurrence of current reversals is analogous.

Let us denote a shift by $m$ lattice sites in the sets of occupation
numbers and site energies by $\bm n_{\to m}$, $\bm\epsilon_{\to m}$
and $\bm\epsilon_{\leftarrow m}$, i.e.\ $(n_{\to m})_i=n_{i-m}$,
$(\epsilon_{\to m})_i=\epsilon_{i-m}$ and $(\epsilon_{\leftarrow
  m})_i=\epsilon_{i+m}$. Because of the $(2m)$-periodicity of the site
energies and the symmetry property
\eqref{eq:eps-inversion-spaceshift}, it holds $\bm\epsilon_{\leftarrow
  m}(t)=\bm\epsilon_{\to m}(t)=-\bm\epsilon(t)$.  We thus obtain,
using Eqs.~\eqref{eq:relation_Gamma+}, \eqref{eq:relation_Gamma-},
\begin{align}
\tilde\Gamma_{i+m}^\pm[\tilde{\bm n}_{\to m},\bm\epsilon(t)]
&=\tilde\Gamma_i^\pm[\tilde{\bm n},\bm\epsilon_{\leftarrow m}(t)]
=\tilde\Gamma_i^\pm[\tilde{\bm n},\bm\epsilon_{\to m}(t)]\nonumber\\[0.5ex]
&=\tilde\Gamma_i^\pm[\tilde{\bm n},-\bm\epsilon(t)]
=\Gamma_i^\pm[\bm n,\bm\epsilon(t)]\,.
\label{eq:Gammarelation-space-shifted}
\end{align}
This means that hopping dynamics in a system with particles and holes
interchanged is the same after a translation in space by $m$.  Given
the path $\mathcal{P}$ in Eq.~\eqref{eq:pathP} with initial condition
$\bm n^{(0)}$, Eq.~\eqref{eq:equal-path-prob} now applies to the
corresponding path $\tilde{\mathcal{P}}=\{\tilde{\bm n}^{(0)}_{\to
  m}(0), \tilde{\bm n}^{(1)}_{\to m}(t_1), \tilde{\bm n}^{(2)}_{\to
  m}(t_2),\ldots\}$.

The equation analogous to Eq.~\eqref{eq:J-relation-nonstationary-1} is
\begin{equation}
J_{i,i+1}(t)\Bigr|_{\mathcal{S}}=\tilde J_{i+m,i+m+1}(t)\Bigr|_{\tilde{\mathcal{S}}}
=-J_{i+m,i+m+1}(t)\Bigr|_{\tilde{\mathcal{S}}}.
\label{eq:J-relation-nonstationary-2}
\end{equation}
Current reversals can thus be identified in space-shifted
nonstationary dynamics when having particle-hole interchanged initial
conditions in two systems with densities $\rho$ and $1-\rho$.

If the time-dependent site energies in
Eq.~\eqref{eq:eps-inversion-spaceshift} drive the system into a
time-periodic steady state, Eq.~\eqref{eq:J-relation-nonstationary-2} implies
for the spatially averaged currents
\begin{align}
\bar J(t,\rho)&=\frac{1}{2m}\sum_{i=1}^{2m}J_{i,i+1}(t)\Bigr|_{\mathcal{S}}
\label{eq:Jrelation-spatialaveraged}
=-\frac{1}{2m}\sum_{i=1}^{2m}J_{i+m,i+m+1}(t)\Bigr|_{\tilde{\mathcal{S}}}\nonumber\\[0.5ex]
&=-\bar J(t,1-\rho)
\end{align}
at any time in the time-periodic steady state. After a further averaging over time,
one obtains $\bar J(\rho)=-\bar J(1-\rho)$.

In the Appendix we derive additional relations between currents
similar to Eqs.~\eqref{eq:J-relation-nonstationary-1} and
\eqref{eq:J-relation-nonstationary-2}, if the site energies
$\epsilon_i(t)$ and $\epsilon_i(t+\tau')$, $\tau'\ge0$, are
axisymmetric or point-symmetric to each other with respect to a
position $x=i$ or $x=i+1/2$ for a lattice site $i$.  These relations
involve complementary sets of occupancy numbers arising from a
mirroring at point $x$.  They allow one to conclude that current
reversal condition (ii) is violated if an axial symmetry is present,
or a point symmetry in addition to Eq.~\eqref{eq:eps-inversion-timeshift}
or Eq.~\eqref{eq:eps-inversion-spaceshift}. In such cases, averaged
particle currents become zero.

\section{Examples of current reversals}
\label{sec:examples}

\subsection{Driven lattice gases}
\label{subsec:current-reversal-examples-driven-lattice-gases}

To demonstrate the general results derived in
Sec.~\ref{sec:reasoning_current_reversal}, we take a lattice gas
Hamiltonian $H_0$ with nearest-neighbor interaction $V$,
i.e.\ $V_\alpha=V\delta_{\alpha1}$ in Eq.~\eqref{eq:H0}.  The energy
differences entering the jump rates in Eqs.~\eqref{eq:DefGamma+},
\eqref{eq:DefGamma-} then are
\begin{align}
\Delta H_{i,i\pm1}[\bm n,\bm\epsilon(t)]=&
V[n_{i\pm2}(n_i-n_{i\pm1})-n_{i\mp1}(n_i-n_{i\pm1})]\nonumber\\
&\hspace{1em}{}+(n_i-n_{i\pm1})[\epsilon_{i\pm1}(t)-\epsilon_i(t)]\,.
\label{eq:DeltaH-example}
\end{align}
When $V=0$, the only interaction is the site exclusion.

For specifying the jump rates, we take the thermal energy as energy
unit and choose
\begin{equation}
f(u)=\nu\exp(-u/2)
\end{equation}
in Eqs.~\eqref{eq:DefGamma+}, \eqref{eq:DefGamma-},
where $\nu$ is an attempt frequency setting our time unit $1/\nu=1$. 
With $\Delta H_{i,i\pm1}$ from Eq.~\eqref{eq:DeltaH-example}, this 
gives
\begin{align}
&\Gamma_i^\pm[\bm n,\bm\epsilon(t)]=\\
&\hspace{1em}n_i\tilde n_{i\pm1}\exp\left\{-\frac{1}{2}\Bigl[V(n_{i\pm2}-n_{i\mp1})
+\epsilon_{i\pm1}(t)-\epsilon_i(t)\Bigr]\right\}.
\nonumber
\end{align}

Site energies $\epsilon_i(t)$ changing sign upon a translation in time
by $\tau/2$ or in space by $m$ are periodic functions also, with
periods $\tau$ or $\lambda=2m$. Traveling waves are thus a generic
type of driving for current reversals to occur. More generally, we can
consider traveling waves with amplitude modulation in space and/or
time:
\begin{equation}
\epsilon_i(t)=A(i,t)w(i-vt)\,,
\label{eq:eps-w}
\end{equation}
Here, $w(x)=w(x+\lambda)$ is a $\lambda$-periodic function specifying
the shape of the wave, $v$ is the wave velocity, and $A(i,t)$ is an
amplitude modulation.  If translations in space and/or time change the
sign of the $\epsilon_i(t)$ [current-reversal condition (i)], and if
the amplitude-modulated wave in Eq.~\eqref{eq:eps-w} generates current
[current-reversal condition (ii)], a current reversal must occur.

We perform kinetic Monte-Carlo simulations to determine currents for
various cases, where we choose a sinusoidal wave
\begin{equation}
w(i-vt)=\sin(\frac{2\pi}{\lambda}(i-vt))
\label{eq:w-sinus}
\end{equation}
as basic wave form.  The simulations are carried out by applying the
first-reaction time algorithm for time-dependent rates
\cite{Holubec/etal:2011}.

\begin{figure}[b!]
\centering
\includegraphics[width=\columnwidth]{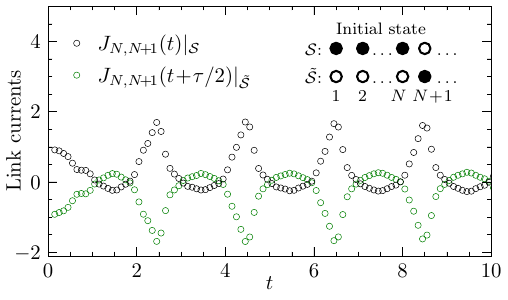}
\caption{Current reversal for nonstationary dynamics in a
  lattice gas under driving by a sinusoidal traveling wave
  [Eq.~\ref{eq:w-sinus}] with $A=3$, $\lambda=4$ and $\tau=2$ and
  site-exclusion interaction ($V=0$), demonstrating
  Eq.~\eqref{eq:J-relation-nonstationary-1} for site energies changing sign under time translation.
  For an initial state in
  a system $\mathcal{S}$ with $N$ particles occupying the first $N$
  lattice sites $i=1,\ldots,N$ (density $\rho=N/L$), and an initial
  state in a system $\tilde{\mathcal{S}}$ with $L-N$ particles
  occupying the last $L-N$ lattice sites $i=N+1,\ldots,L$ (density
  $1-\rho$), the link current $J_{N,N+1}(t+\tau/2)$ in system
  $\tilde{\mathcal{S}}$ is the negative of the link current
  $J_{N,N+1}(t)$ in system $\mathcal{S}$.}
\label{fig:link_currents_nonstationary}
\end{figure}

As a first example, we consider the current reversal relation
\eqref{eq:J-relation-nonstationary-1} in nonstationary dynamics for a
constant $A$ in Eq.~\eqref{eq:eps-w}.  In this case, the wave form has
a glide-reflection symmetry under a translation by $\lambda/2$:
\begin{equation}
w\left(x+\frac{\lambda}{2}\right)=w\left(x+\frac{v\tau}{2}\right)=-w(x)\,.
\label{eq:w-symmetry}
\end{equation}
Both Eqs.~\eqref{eq:eps-inversion-timeshift} and
\eqref{eq:eps-inversion-spaceshift} are fulfilled in this case.

For the initial state $\bm n^{(0)}$ in system $\mathcal{S}$ at time
$t=0$, we take all sites $i=1,\ldots,N$ to be occupied and all other
to be empty, i.e.\ $n_i^{(0)}=1$ for $i=1,\ldots,N$, and $n_i^{(0)}=0$
for $i=N+1,\ldots,L$. This initial state and the particle-hole
inverted one in system $\tilde{\mathcal{S}}$ at time $t=\tau/2$ are
illustrated in the upper right corner of
Fig.~\ref{fig:link_currents_nonstationary}. As shown in the figure,
the simulated currents for the link $(N,N+1)$ satisfy relation
\eqref{eq:J-relation-nonstationary-1}.

\begin{figure}[t!]
\centering
\includegraphics[width=\columnwidth]{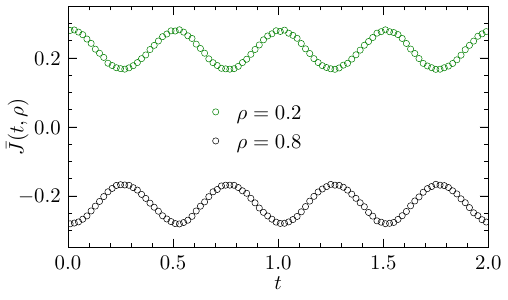}
\caption{Current reversal for time-dependent spatially averaged
  currents in the time-periodic steady state for the same driving as in
  Fig.~\ref{fig:link_currents_nonstationary}, demonstrating
  Eq.~\eqref{eq:Jrelation-spatialaveraged} for site energies changing sign under spatial translation.  
  When averaging over one
  spatial period $\lambda$ of the sinusoidal traveling wave in
  Eq.~\eqref{eq:w-sinus}, $\bar J(t,\rho)=-\bar J(t,1-\rho)$ in
  systems with densities $\rho=0.2$ and $\rho=0.8$.}
\label{fig:spatially_averaged_currents}
\end{figure}

In the second example, we check the relation
\eqref{eq:Jrelation-spatialaveraged} for spatially
averaged currents in the time-periodic steady state for the same
traveling-wave driving with constant amplitude.  Spatially averaged
currents at particle densities $\rho=0.2$ and $\rho=0.8$ are compared
in Fig.~\ref{fig:spatially_averaged_currents}.  At all times $t$, one
current is the negative of the other, $\bar J(t,\rho)=-\bar
J(t,1-\rho)$.

\begin{figure}[t!]
\centering
\includegraphics[width=\columnwidth]{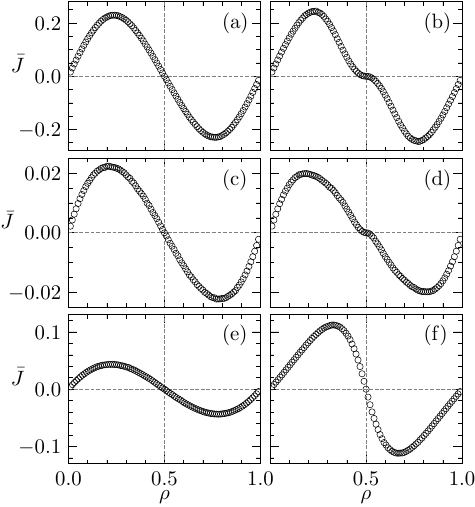}
\caption{Demonstration of reversals in time-averaged currents
of lattice gases according to Eq.~\eqref{eq:Jrelation-time-averaged}
under driving by a sinusoidal traveling wave for (a),
  (b) constant amplitude $A=3$, (c), (d) time-modulated amplitude
  given in Eq.~\eqref{eq:Amod-time}, and (e), (f) superposition of two
  traveling waves with spatially modulated amplitudes given in
  Eq.~\eqref{eq:two-wave-driving}. 
  In the left panels (a), (c), and
  (e), particles interact with site exclusion only ($V=0$), while in
  the right panels (b), (d), and (f), an additional strong repulsive
  interaction $V=10$ is present.  The wavelength $\lambda=4$ and time
  period $\tau=2$ in (a)-(d) are the same as in
  Fig.~\ref{fig:link_currents_nonstationary}.  In (e) and (f),
  $\lambda=8$ and $\tau=2$.}
\label{fig:current_reversals_DLG}
\end{figure}

In Fig.~\ref{fig:current_reversals_DLG}, we demonstrate reversals
$\bar J(\rho)=-\bar J(1-\rho)$ of time-averaged currents in time-periodic
steady states.  In the left column of the graphs
[Figs.~\ref{fig:current_reversals_DLG}(a), (c), (e)], results are
displayed for $V=0$ (only site-exclusion interaction), and in the
right column [Figs.~\ref{fig:current_reversals_DLG}(b), (d), (f)]
results for $V=10$ (additional strong repulsive nearest-neighbor
interaction).

In Figs.~\ref{fig:current_reversals_DLG}(a) and (b), the results are
for constant amplitude in Eq.~\eqref{eq:eps-w}, while in
Figs.~\ref{fig:current_reversals_DLG}(c) and (d) the amplitude of the
traveling wave is modulated in time as
\begin{equation}
A(t)=3\sin(\frac{2\pi t}{\tau})\,.
\label{eq:Amod-time}
\end{equation}

Possibilities to realize the current-reversal conditions (i) and (ii)
are quite rich. In Figs.~\ref{fig:current_reversals_DLG}(e) and (f) we
have taken a superposition of two traveling waves with spatially
modulated amplitudes as driving,
\begin{align}
\epsilon_i(t)&=3\sin(\frac{2\pi i}{\lambda})
\left[\sin(\frac{2\pi i}{\lambda})\cos(\frac{2\pi}{\lambda}(i-vt))\right.\nonumber\\
&\left.\hspace{7em}{}
+\cos(\frac{2\pi i}{\lambda})\sin(\frac{2\pi}{\lambda}(i-vt))\right]\nonumber\\
&=3\sin(\frac{2\pi i}{\lambda})\sin(\frac{2\pi}{\lambda}(2i-vt))\,.
\label{eq:two-wave-driving}
\end{align}

Current reversals with $\bar J(\rho)=-\bar J(1-\rho)$ occur if the
conditions (i) and (ii) of Sec.~\ref{sec:reasoning_current_reversal}
are fulfilled. If these conditions are not met, but can be recovered
by a suitable choice of parameters in a model, one can expect current
reversals to be present without point symmetry.  An example is the
driving of lattice gases by a traveling wave plus a constant bias
\cite{Dierl/etal:2014}.

\subsection{Driven Brownian motion}
\label{subsec:current-reversal-Brownian}
We finally consider the question whether current reversals are also
possible for driven Brownian motion of interacting particles. While
driven lattice gases are particularly interesting for studying
fundamental questions of nonequilibrium physics, driven Brownian
motion has a wider range of applications, including tailored
experiments, e.g., with colloidal particles \cite{Wei/etal:2000,
  Lutz/etal:2004a, Tierno:2014, Juniper/etal:2015, Juniper/etal:2016,
  Juniper/etal:2017, Illien/etal:2017, Berner/etal:2018,
  Mirzaee-Kakhki/etal:2020a, Lips/etal:2021,
  Villada-Balbuena/etal:2021, Cereceda-Lopez/etal:2021,
  Leyva/etal:2022, Cereceda-Lopez/etal:2023}.

A model of Brownian motion reflecting driven lattice gas dynamics in
certain parameter ranges is the Brownian asymmetric simple exclusion
process (BASEP) \cite{Lips/etal:2018, Lips/etal:2019, Lips/etal:2020}.
In the BASEP, hard spheres with diameter $\sigma$ are driven across a
periodic potential. For a constant drag force and sinusoidal potential
of wavelength $\lambda$, particle diameters in the range
$0.7-0.85\lambda$ and particle number densities below
$\rho=1/\lambda$, current-density relations and particle dynamics on a
coarse-grained scale closely resemble that of the asymmetric simple
exclusion process \cite{Derrida:1998, Schuetz:2001, Chou/etal:2011,
  Mallick:2015}, that is a lattice gas with site-exclusion interaction
and biased jumps in one direction.

\begin{figure}[b!]
\centering
\includegraphics[width=\columnwidth]{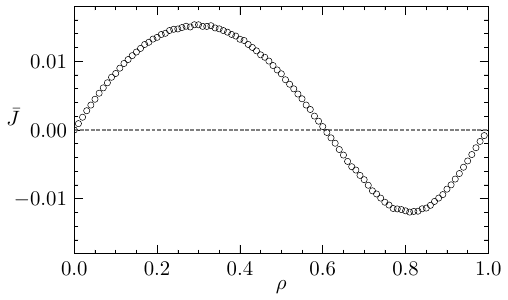}
\caption{Current reversal in one-dimensional Brownian motion of
  hard spheres with diameter $\sigma$ in a sinusoidal potential with
  amplitude $U_0/2$ and under driving by a sinusoidal traveling wave
  with same amplitude, according to the time-dependent potential in
  Eq.~\eqref{eq:U-Brownian}.  Parameters are $\sigma=\lambda=1$,
  $U_0=6k_{\rm B}T$, and $v=0.2\, k_{\rm B}T\mu/\lambda$ ($\mu$:
  particle mobility).}
\label{fig:current_reversal_BASEP}
\end{figure}

If the periodic potential in the BASEP is a traveling wave, the
Brownian dynamics in a frame comoving with the wave is equal to that
in the periodic potential under a constant drag force. Due to this
fact, one can show that current reversals do not occur, because they
would lead to negative entropy production in the time-periodic steady state
\cite{Lips/etal:2019}.  We can expect them to emerge, however, when a
traveling wave is driving the particles across a periodic energy
landscape.

To test this, we have performed Brownian simulations for the BASEP
with the time-dependent external potential
\begin{equation}
U(x,t)=\frac{U_0}{2}\left[\cos(\frac{2\pi x}{\lambda})+\sin(\frac{2\pi}{4\lambda} (x-vt))\right]\,.
\label{eq:U-Brownian}
\end{equation}
The Langevin equations of motion under consideration of the
hard-sphere constraints are solved by applying the algorithm developed
in Refs.~\cite{Antonov/etal:2022c, Antonov/etal:2025a}. As length unit
we choose the wavelength $\lambda$, and as time unit $\lambda^2/k_{\rm
  B}T\mu$, where $\mu$ is the particle mobility.  Results for the
simulated current in the time-periodic steady state are shown in
Fig.~\ref{fig:current_reversal_BASEP} for hard-sphere diameter
$\sigma=\lambda=1$, $U_0/k_{\rm B}T=6$, and $v=1/5$. In agreement
with our conjecture, a current reversal occurs from positive to
negative current at $\rho\cong0.605$.

\section{Conclusions}
\label{sec:conclusions}
We have shown that current reversals with symmetry 
$\bar J(\rho)=-\bar J(1-\rho)$ emerge generally in lattice gases driven by a
time-dependent potential that changes sign upon a translation in time
and/or space. This result is valid for arbitrary pair interactions. 
Examples were given for short-range 
repulsive nearest-neighbor interactions. For longer-range or attractive interactions,
clustering effects can lead to long transients before stationary states are reached.
Current reversals occur also for nonstationary dynamics for particle-hole inverted
initial conditions and to time-dependent spatially-averaged currents
in stationary states. The condition of sign inversion is in particular
fulfilled for traveling-wave potentials, including ones with temporal
and/or spatial amplitude modulation. For various examples of such
traveling waves, we have demonstrated the occurrence of current
reversals by kinetic Monte Carlo simulations.

The reason for the current reversals is that holes see the energy 
landscape inverted, while experiencing the same interaction as the particles. 
Hence, when the energy landscape inverts under a translation in time or space, 
the hole dynamics becomes the same as that of particles in a system with density $1-\rho$. 
Since particle currents are the negative of hole currents, one finds $\bar J(\rho)=-\bar J(1-\rho)$. 

Without resorting to symmetry considerations, one can offer the following intuitive
argument for why currents in lattice gases generally flow opposite to the driving at high particle densities:
Holes prefer to occupy sites with high energy. When
considering the limit $\rho\to1$, i.e.\ small hole densities, they tend to be
dragged with the driving at their favorable positions without being perturbed by other holes. 
Accordingly, a hole current results in the driving direction, corresponding to a particle current against it.

When weakly breaking the condition of sign inversion of the potential
under time- or space-translation,
the relation $\bar J(\rho)=-\bar J(1-\rho)$
is no longer perfectly obeyed.
Current reversals still occur, because current-density relations change continuously with the potential, 
except for possible singularities at phase transitions. However, the point of current inversion 
shifts away from the particle density $\rho=1/2$. 
Typically, an asymmetry of current amplitudes
arises for particle flows in the direction of and opposite to the driving.
An example is given in Ref.~\cite{Dierl/etal:2014},
where a constant bias is added to a traveling-wave potential, 
causing it to violate the symmetry requirements,
i.e.\ not to invert sign under a temporal or spatial translation.

While we have presented our derivations for one dimensional lattice
gases, the reasoning based on particle-hole symmetry are equally valid
in higher dimensions.  Our general result is useful also to understand
and conjecture current reversals if the driving potential has no ideal
translation symmetry but can be viewed as perturbed version of such
potential.  Current-density relations will be asymmetric then in
general.

One furthermore should expect current reversals upon a change of
particle density to appear in continuous-space dynamics of interacting
particles when these resemble driven lattice gas dynamics on a
coarse-grained scale.  As an example, we showed this for Brownian
motion of hard spheres driven across a periodic energy landscape by a
traveling wave.  This example can be directly realized in experiments
with colloidal particles \cite{Cereceda-Lopez/etal:2021, Cereceda-Lopez/etal:2023}.

\appendix*

\section{Implications of axisymmetric or point-symmetric site energies on particle currents}
\label{app:J-relations-for-addition-eps-symmetries}
A further symmetry operation is a mirroring of occupation numbers $\bm
n$ and site energies $\bm\epsilon(t)$ at a position $x=i$ or $x=i+1/2$
for a lattice site $i$.  Denoting correspondingly mirrored sets by
$\hat{\bm n}$, and $\hat{\bm\epsilon}(t)$, i.e.\
\begin{align}
\hat n_i&=n_{2x-i}\,,\\
\hat\epsilon_i(t)&=\epsilon_{2x-i}(t)\,,
\end{align}
the jump rates in the corresponding configurations satisfy
\begin{equation}
\Gamma_i^\pm[\bm n,\bm\epsilon(t)]=\Gamma_{2x-i}^\mp[\hat{\bm n},\hat{\bm\epsilon}(t)]\,.
\end{equation}
This can be formally proven by resorting to Eqs.~\eqref{eq:DefGamma+},
\eqref{eq:DefGamma-}, but is immediately evident also from the fact
that the mirroring at point $x$ corresponds to a spatial inversion of
the hopping dynamics.

\vspace*{2ex}
\centerline{\it Axisymmetric site energies}
\vspace{0.5ex} 
Site energies $\epsilon_i(t)$ and
$\epsilon_i(t+\tau')$, $\tau'\ge0$, being axisymmetric with respect to
$x$ satisfy
\begin{equation}
\epsilon_{2x-i}(t+\tau')=\epsilon_i(t)\,.
\end{equation}
By replacing $i$ with $2x-i$, this symmetry relation can also be
written in the form $\epsilon_{2x-i}(t)=\epsilon_i(t+\tau')$.
Accordingly,
\begin{equation}
\Gamma_i^\pm[\bm n,\bm\epsilon(t)]
=\Gamma_{2x-i}^\mp[\hat{\bm n},\hat{\bm\epsilon}(t)]
=\Gamma_{2x-i}^\mp[\hat{\bm n},\bm\epsilon(t+\tau')]\,.
\end{equation}
This means that the hopping dynamics at time $t$ is equal to that in
reversed direction for the system with mirrored occupation numbers at
time $t+\tau'$.

Given the path in Eq.~\eqref{eq:pathP} with initial state $\bm
n^{(0)}$, the path
\begin{equation}
\hat{\mathcal{P}}=\{\hat{\bm n}^{(0)}(\tau'), \hat{\bm n}^{(1)}(t_1+\tau'), \hat{\bm n}^{(2)}(t_2+\tau'),\ldots\}
\label{eq:pathhatP}
\end{equation}
with mirrored states at times $\tau', t_1+\tau',\ldots$ has the same
probability.  Averaging over all corresponding paths for given initial
states yields
\begin{align}
J_{i,i+1}(t)\Bigr|_{\bm n^{(0)}}
&=\langle \Gamma_i^+[\bm n,\bm\epsilon(t)]
-\Gamma_{i+1}^-[\bm n,\bm\epsilon(t)]\rangle_{\bm n^{(0)}}\nonumber\\
&\hspace{-3.75em}=\langle \Gamma_{2x-i}^-[\hat{\bm n},\bm\epsilon(t+\tau')]-
\Gamma_{2x-i-1}^+[\hat{\bm n},\bm\epsilon(t+\tau')]\rangle_{\hat{\bm n}^{(0)}(\tau')}
\nonumber\\
&=-J_{2x-i-1,2x-i}(t+\tau')\Bigr|_{\hat{\bm n}^{(0)}(\tau')}\,.
\end{align}
For nonstationary dynamics with initial state $\bm n^{(0)}$, this
means that the current $J_{i,i+1}(t)$ is equal to the negative of the
current $J_{2x-i-1,2x-i}(t+\tau')$ with initial state $\hat{\bm
  n}^{(0)}(\tau')$ at time $\tau'$.

When taking an average over all lattice sites, we can write
\begin{equation}
\bar J(t,\rho)=-\bar J(t+\tau',\rho)
\end{equation}
If the time-dependent $\epsilon_i(t)$ drive the system into a
time-periodic stationary state, $\bar J(\rho)=-\bar J(\rho)$ after
averaging over a time period. Hence, $\bar J(\rho)=0$ in such case.

\vspace*{2ex}
\centerline{\it Point-symmetric site energies}
\vspace{0.5ex}
Site energies $\epsilon_i(t)$ and $\epsilon_i(t+\tau')$, $\tau'\ge0$,
being point-symmetric with respect to $x$ satisfy
\begin{equation}
\epsilon_{2x-i}(t+\tau')=-\epsilon_i(t)\,.
\label{eq:point-symmetric-epsilon}
\end{equation}
By replacing $i$ with $2x-i$, this symmetry relation can also be
written in the form $\epsilon_{2x-i}(t)=-\epsilon_i(t+\tau')$.  Using
Eqs.~\eqref{eq:relation_Gamma+}, \eqref{eq:relation_Gamma-}, we find
\begin{align}
\Gamma_i^\pm[\bm n,\bm\epsilon(t)]
&\!=\!\Gamma_{2x-i}^\mp[\hat{\bm n},\hat{\bm\epsilon}(t)]
\!=\!\Gamma_{2x-i}^\mp[\hat{\bm n},-\bm\epsilon(t\!+\!\tau')]
\label{eq:Gamma-relation-point-symmetric-epsilon}\\
&=\tilde\Gamma_{2x-i}^\mp[\tilde{\hat{\bm n}},\bm\epsilon(t+\tau')]
=\Gamma_{2x-i\mp1}^\pm[\tilde{\hat{\bm n}},\bm\epsilon(t+\tau')]
\nonumber\,.
\end{align}
Here, $\tilde{\hat{\bm n}}$ is the state obtained after mirroring
state $\bm n$ and additional particle-hole exchange.
Equation~\eqref{eq:Gamma-relation-point-symmetric-epsilon} means that
the hopping dynamics at time $t$ is equal to that in a system with
particle-hole exchanged and mirrored occupation numbers at time
$t+\tau'$.

Given the path in Eq.~\eqref{eq:pathP} in a system $\mathcal{S}$ with
initial state $\bm n^{(0)}$, the path
\begin{equation}
\tilde{\hat{\mathcal{P}}}
=\{\tilde{\hat{\bm n}}^{(0)}(\tau'), \tilde{\hat{\bm n}}^{(1)}(t_1+\tau'), \tilde{\hat{\bm n}}^{(2)}(t_2+\tau'),\ldots\}
\label{eq:pathP-tildehatP}
\end{equation}
with initial state $\tilde{\hat{\bm n}}^{(0)}(\tau')$ at time $\tau'$
in system $\tilde{\mathcal{S}}$ with exchanged particle and holes has
the same probability.  Accordingly,
\begin{align}
J_{i,i+1}(t)\Bigr|_\mathcal{S}&=\langle \Gamma_i^+[\bm n,\bm\epsilon(t)]
-\Gamma_{i+1}^-[\bm n,\bm\epsilon(t)]\rangle_{\scriptscriptstyle\mathcal{S}}\nonumber\\
&=\langle \Gamma_{2x-i-1}^+[\tilde{\hat{\bm n}},\bm\epsilon(t+\tau')]
-\Gamma_{2x-i}^-[\tilde{\hat{\bm n}},\bm\epsilon(t+\tau')]\rangle_{\scriptscriptstyle\tilde{\mathcal{S}}}
\nonumber\\
&=J_{2x-i-1,2x-i}(t+\tau')\Bigr|_{\tilde{\mathcal{S}}}\,.
\label{eq:Jlink-relation-point-symmetric-epsilon}
\end{align}
For nonstationary dynamics of a system $\mathcal{S}$ with initial
state $\bm n^{(0)}$, this means that the current $J_{i,i+1}(t)$ is
equal to the current $J_{2x-i-1,2x-i}(t+\tau')$ in a system
$\tilde{\mathcal{S}}$ with mirrored and particle-hole exchanged
initial state $\tilde{\hat{\bm n}}^{(0)}$ at time $\tau'$.

We note that it does not follow $J_{i,i+1}(t)=J_{i,i+1}(t+2\tau')$
from Eq.~\eqref{eq:Jlink-relation-point-symmetric-epsilon} in general,
because this equation holds for a given initial state, while for
relating currents at times $t+\tau'$ and $t+2\tau'$ one would need to
consider the ensemble of initial states following from the time
evolution between times $t$ and $t+\tau'$.  For stationary states,
where initial states are irrelevant, one can conclude
$J_{i,i+1}(t)=J_{i,i+1}(t+2\tau')$.

When taking an average over all lattice sites,
\begin{equation}
\bar J(t,\rho)=\bar J(t+\tau',1-\rho)\,.
\label{eq:J(t,rho)-relation-point-symmetric-epsilon}
\end{equation}
If the time-dependent $\epsilon_i(t)$ drive the system into a
time-periodic stationary state $\bar J(\rho)=\bar J(1-\rho)$ after
averaging over a time period.

If in addition to the
point-symmetry~\eqref{eq:point-symmetric-epsilon}, the site energies
satisfy Eq.~\eqref{eq:eps-inversion-timeshift} or Eq.~\eqref{eq:eps-inversion-spaceshift}, 
i.e.\ invert sign
after time translation by $\tau/2$, and one of the two times are
commensurate such that $p\tau=q\tau'$ for coprime $p,q\in\mathbb{N}$,
a time-averaging over $p\tau=q\tau'$ in the stationary state yields
$\bar J(\rho)=\bar J(1-\rho)=-\bar J(1-\rho)$, i.e.\ $\bar J(\rho)=0$.
If $\tau/\tau'$ is irrational, $\bar J(\rho)\to0$ in the stationary
state with increasing averaging time.

\vspace{2ex}
\begin{acknowledgments}
This work has been funded by the Deutsche Forschungsgemeinschaft (DFG,
Project No.\ 355031190). We sincerely thank A.~Ryabov and the members
of the DFG Research Unit FOR2692 for fruitful discussions. We
acknowledge use of a high-performance computing cluster funded by the
Deutsche Forschungsgemeinschaft (DFG, Project No.\ 456666331).
\end{acknowledgments}


%

\end{document}